\documentclass[%
 reprint,
 amsmath,amssymb,
 aps,
prb,
]{revtex4-2}
\usepackage{graphicx}
\usepackage{dcolumn}
\usepackage{bm}

\begin{document}

\preprint{APS/123-QED}

\title{Constructive and Destructive Interference 
of Kerker-type Scattering in an Ultra-thin 
Silicon Huygens Metasurface}

\author{Xia Zhang}
\author{Jing Li}
\author{John F. Donegan}
\author{A. Louise Bradley}
 \email{bradlel@tcd.ie}

\affiliation{%
School of Physics, CRANN and Amber, Trinity College Dublin
}%

\date{\today}

\begin{abstract}
High refractive index dielectric nanoparticles have provided a new platform for exotic light manipulation through the interference of multipole modes. The Kerker effect is one example of a Huygens source design. Rather than exploiting interference between the electric dipole and magnetic dipole, as in many conventional Huygens source designs, we explore Kerker-type suppressed backward scattering mediated by the dominant electric dipole, toroidal dipole and magnetic quadrupole. These modes are provided by a designed and fabricated CMOS compatible ultra-thin Silicon nanodisk metasurface with a suppressed magnetic dipole contribution, and verified through multipole decomposition. The non-trivial substrate effect is considered using a semi-analytical transfer matrix model. The model successfully predicts the observed reflection dip. By applying a general criterion for constructive and destructive interference, it is shown that while constructive interference occurs between the electric and toroidal dipole contributions, the experimentally observed suppressed backward Kerker-type scattering arises from the destructive interference between backward scattered contributions due to the total electric dipole and the magnetic quadrupole. Our study paves the way towards new types of Huygens sources or metasurface design, such as for peculiar transverse Kerker scattering. 

\end{abstract}

\maketitle


\section{Introduction and Motivation}

Huygens' Principle states that all points of a wave front of light can be regarded as a new source of secondary waves that expand in every direction. Metasurfaces encompass an array of elements, each of which behaves as secondary sources, providing a platform for control of the radiation \cite{holloway2012overview, pfeiffer2013metamaterial, yu2014flat, chen2016review}. The light transmitted, reflected or refracted through a metasurface can be manipulated to obtain a desired direction, wavefront, phase and polarization. This control is realized by designing the nanoresonator shape, size, material, as well as their geometrical arrangement in one unit cell and period  \cite{ding2017gradient, babicheva2017reflection}. A dielectric metasurface can have much higher efficiency compared with a plasmonic counterpart, which suffer from ohmic losses. Additionally, a dielectric metasurface, or an array of dielectric nanoresonators, sustains overlapping resonant multipolar resonances  \cite{kuznetsov2016optically, staude2017metamaterial, babicheva2017resonant}. The interplay of multipolar resonances in a metasurface or a single nanoresonator provides another degree of freedom for light tailoring, such as the Kerker effect, the anti-Kerker  effect \cite{kerker1983electromagnetic,geffrin2012magnetic, terekhov2017multipolar}, and transverse  Kerker scattering \cite{bag2018transverse, shamkhi2019transverse, liu2019lattice}. Kerker-type scattering, or highly suppressed backward scattering, attracts much interest as it enables a concentrated electric field inside the dielectric structure and far-field scattering in the forward direction. Multipole decomposition is a useful approach to gain insight into the contribution of each individual multipole mode as well as the total contributions determining the far-field scattering of a single nanoresonator  \cite{evlyukhin2016optical} or a metasurface \cite{staude2013tailoring}. Currently, most of the theoretical work considers dielectric metasurfaces in air medium  \cite{terekhov2017multipolar,terekhov2017resonant, liu2018generalized} and only a few theoretical work aims to explore the scattering properties taking account of the effect of the  substrate  \cite{babicheva2017reflection,berkhout2020simple}. We fill this gap experimentally, numerically and semi-analytically using a dielectric metasurface fabricated from a common silicon-on-insulator wafer. Our study on the effect of substrate may guide the scattering application in photonic nanojets lithography \cite{kallepalli2013long} or photonic nanojets  design \cite{pacheco2019photonic}.

Conventionally, scattering manipulation exploits the interference between the electric dipole (ED) moment and magnetic dipole (MD) moment  \cite{decker2015high,alaee2015generalized, babicheva2017reflection,babicheva2017resonant, alaee2015generalized}. Recently, interference between the ED and  toroidal  dipole (TD) has  attracted  great  interest. The electric dipole in Cartesian coordinate is expressed as $\textbf{p}=\frac{i}{\omega}\int \textbf{j}d\textbf{r}$, where $\textbf{j}=\epsilon_0(\epsilon_{Si}-1)\textbf{E(\textbf{r})}$ is the induced polarization current inside the nanoparticle. The toroidal dipole moment is
$\textbf{T}=-\frac{i\omega}{10c}\int[(\textbf{r}\cdot\textbf{j}(\textbf{r}))\textbf{r}-2\textbf{r}^2\textbf{j}(\textbf{r})]d\textbf{r}$. The coherent sum of the ED and TD, $\textbf{p}+ik\textbf{T}$, known as the total electric dipole \cite{terekhov2017multipolar} is denoted as TED. The TED can be invisible or non-radiative when $\textbf{p}=-ik\textbf{T}$, referred to as an “anapole mode”  \cite{terekhov2019multipole, wu2018optical, miroshnichenko2015nonradiating}. A recent review on the TD and how it can be used for exotic light manipulation can be found in reference \cite{gupta2020toroidal}. An ideal non-radiating anapole mode arises from the complete destructive interference of the radiation from the ED and the TD \cite{wei2016excitation,takou2019dynamic,parker2020excitation}. 
There have been many experimental reports of anapole modes identified through their near-field distribution of opposite closed loop circulating current densities \cite{grinblat2016enhanced}. In the broader context, it is defined as low-radiation or suppressed scattering \cite{baryshnikova2019optical} arising from partial destructive interference.  However, it is also found, by tuning the geometrical factors in high refractive index cylindrical nanostructures, that constructive interference between the scattering contributions from the ED and TD can be achieved, termed the super-dipole state \cite{terekhov2017multipolar}.  It is worth noting that an ED can also manifest as a dip in far-field scattering profile of a single nanodisk \cite{wang2016engineering} or a single nanowire  \cite{liu2015invisible}. Additionally, under most circumstances where the ED moment and TD moment do not generate scatter fields of comparable amplitude, it is difficult to conclude from the far-field radiation pattern how the scattered contributions interfere with each other. This is the same principle as the visibility of interference, where a clear visible bright and dark pattern, or ideal unity visibility, requires that the interfering waves share the same amplitude and simultaneously meet the phase relationship. If a disparity of amplitude exists between the interfering two waves, the interference is more difficult to discern. Therefore, the experimentally observed near-field scattering pattern with a circulating electric field profile and/or a suppressed far-field scattering is insufficient to conclude on the existence of super-dipole or anapole modes. Multipole decomposition is thus pivotal to understanding the mechanisms that produce the far-field scattered power.

The amplitude and phase of the electromagnetic radiation generated by each of the multipole contributions determines the observed far-field scattered power. Once the modes can be identified, the question arises as to whether the scattered radiation contributions interfere constructively or destructively. One method that is used is to compare  the amplitude of the ED/TD and TED contributions \cite{terekhov2019multipole,terekhov2017resonant}. This can clearly demonstrate the destructive interference mechanisms but is insufficient to conclude constructive interference. If the amplitude of the coherent sum of the ED and TD amplitudes, TED, is smaller than ED or TD, it can be concluded that destructive interference exists. However, constructive interference cannot be concluded from an increase of the TED amplitude compared to ED or TD amplitudes. Here we present an approach to identify constructive and destructive interference through analysis of the scattered power of the multipole modes. Additionally, in contrast to the most studies of interference arising due to the ED and MD, we performed our study on a specifically designed ultra-thin Si metasurface with suppressed MD contribution but ED, TD and magnetic quadruple (MQ)  mode so that we can explore the interference between their scattered contributions. We also include the effects of the substrate on the measured optical properties of the metasurface.

\section{Experiment, Simulation and Model}

Silicon (Si) nanodisk metasurfaces are fabricated on a Si-on-insulator (SOI) wafers (45 nm top Si thickness, 150 nm buried oxide thickness, Soitec). Electron-
 lithography is performed on spin-coated PMMA 950 layer (A3, 3000 rpm) by Electron beam (Elionix ELS 7700), followed by the development in MIBK:IPA 1:3. A 20 nm thickness of chromium (Cr) layer was deposited by e-beam evaporator (Temescal). After lift-off process in hot Remover 1165, a Cr layer was used as a hardmask, the pattern was transferred to the SOI substrate  by Inductively Coupled Plasma Etching (ICP) through the top Si layer with an over-etched thickness of (10$\pm$3) nm BOX layer, verified by Ellipsometry. The thickness of SiO2 is 140 nm. The Cr mask was finally removed using a commercial Cr etchant from Sigma.

\begin{figure}
\centering
\includegraphics[width=\linewidth]{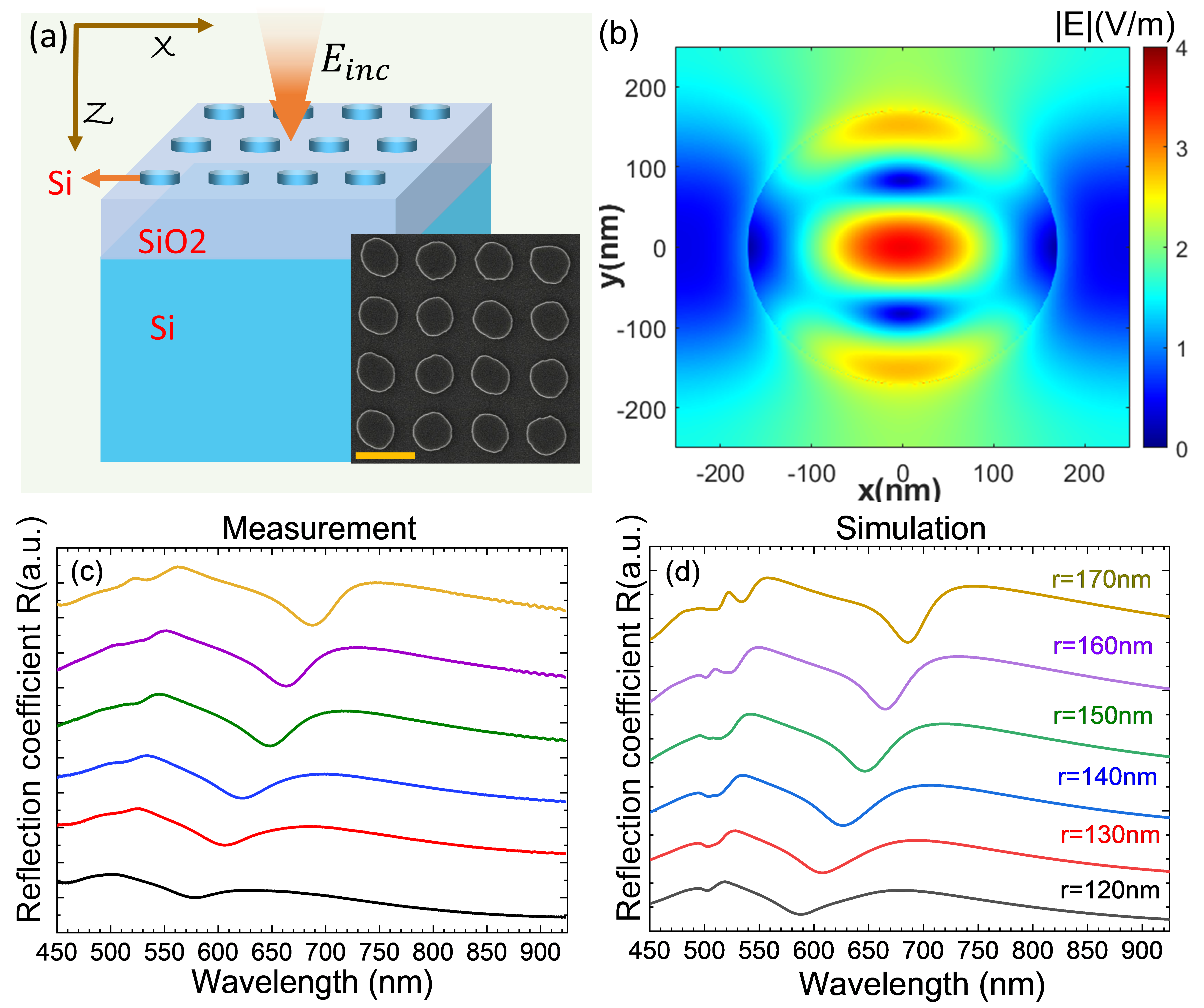}
\caption{(a) Schematic of the periodic Si nanodisk metasurface. The thickness of SiO2 is 140 nm. The scanning electron microscopy image of one metasurface is shown in the inset, with a scale bar corresponding to 500 nm. A normally incident plane wave propagating in the $z$ direction and polarized along the $x$-axis, $\mathrm {E_{inc}=E_0}e^{i(kz-\omega t)}\hat{\textbf{x}}$, interacts with the metasurface. (b) Electric field map within in one period for a nanodisk with r = 170 nm at a wavelength of 686 nm corresponding to the reflection dip wavelength 686 nm. The field map has an anapole-like pattern. (c) The measured and (d) numerically simulated reflection coefficient of the Si nanodisk metasurfaces with nanodisk radii of 120 nm to 170 nm.}
\label{fig:F1}
\end{figure}

In order to achieve a metasurface with a suppressed MD and dominant TED and MQ contributions, arrays of ultra-thin Si nanodisks with a height of 45 nm and radii varying between 120 nm to 170 nm are fabricated. The thickness of SiO2 is 140 nm. The period in the $x$ and $y$ directions is 500 nm. Each metasurface has an area of 50 $\mu $m$\times$50   $\mu$m. A scanning electron microscopy image of one metasurface is shown in the inset of Fig. \ref{fig:F1}(a). To measure the reflectance spectrum a white light beam passes through the objective lens (5x,  NA=0.12) and is incident on the  sample surface at near- normal incidence, within an angle of 7$^0$. The illumination diameter is approximately 6 $\mu$m.  The reflected light is collected by the same objective and coupled to a CCD camera and spectrometer. The reflected signal is normalized by the signal reflected from a silver mirror. A schematic of the simulation under illumination is shown in Fig \ref{fig:F1}(a). The total electric field distribution in one Si nanodisk is computed using a three-dimensional finite-difference time-domain (FDTD) method (Lumerical, Inc). The incident wave is linearly polarized along the $x$ axis and the wavevector $k$ is along the $z$ axis. The wavelength-dependent real and imaginary parts of refractive index and permittivity is obtained from fitting the experimental data in reference \cite{aspnes1983dielectric}.  The simulation area is 500$\times$500$\times$6000 nm$^3$ and the mesh size is 2.5 nm. The simulation is performed with periodic boundary conditions in the $x$ and $y$ directions and the perfectly matched layers in the $z$ direction.

As seen in Fig. \ref{fig:F1}(c) there is a clear dip in the experimentally observed reflection spectra, that shifts to higher wavelengths with increasing nanodisk radius. The reflection dip indicates the suppressed backward scattering or Kerker-type scattering. Fig. \ref{fig:F1}(d) shows excellent agreement of the numerically simulated spectra with the experimental data. Slight broadening of the measured spectra is observed due to inhomogeneously broadening from fabrication imperfection and non-perfect normal incidence in the measurement. As the disk radius is varied from 120 nm to 170 nm, the aspect ratio, defined as the ratio of height to diameter, is changed between 0.13 to 0.18. In contrast to most work on manipulating interference between the ED and MD, these ultra-thin metasurfaces are the key to obtaining a suppressed MD and dominant ED, TD and MQ modes.  The electric field distribution shown in Fig. \ref{fig:F1}(b) for a metasurface with a nanodisk of radius 170 nm indicates interference arising from the ED and TD mode contributions \cite{fu2013directional,wang2016engineering,li2018origin}. As stated earlier, excitation of the ED-alone can also manifest a dip in scattered radiation profile  \cite{wang2016engineering}.  In order to explore the presence of super-dipole or anapole modes and to gain further insight into the nature of the interfering modes, multipole decomposition is performed.

According to the standard expansion method, we consider the decomposition of the multipolar modes up to quadrupole modes by integrating over the nanodisk volume \cite{terekhov2017multipolar}.
\begin{equation}
\begin{split}
&\textbf{p}=\int \epsilon_0(\epsilon_{Si}-1)\textbf{E}(\textbf{r})d\textbf{r}
\\&\textbf{T}=\frac{-i\omega}{10c}\int\epsilon_0(\epsilon_{Si}-1)[(\textbf{r}\cdot\textbf{E}(\textbf{r}))\textbf{r}-2\textbf{r}^2\textbf{E}(\textbf{r})]d\textbf{r}
\\&\textbf{m}=-\frac{i\omega}{2}\int\epsilon_0(\epsilon_{Si}-1)[\textbf{r}\times\textbf{E}(\textbf{r})]d\textbf{r}
\\&\textbf{Q}=3\int\epsilon_0(\epsilon_{Si}-1) [\textbf{r}\textbf{E}(\textbf{r})+\textbf{E}(\textbf{r})\textbf{r}-\frac{2}{3}(\textbf{r}\cdot\textbf{E}(\textbf{r}))\hat{U}]d\textbf{r}
\\&\textbf{M}=\frac{\omega}{3i}\int\epsilon_0(\epsilon_{Si}-1)[(\textbf{r}\times\textbf{E}(\textbf{r}))\textbf{r}+\textbf{r}(\textbf{r}\times\textbf{E}(\textbf{r}))]d\textbf{r}
\end{split}
\end{equation}
where $\textbf{r}$ is the coordinate vector with its origin placed at the center of nanodisk. $\textbf{E}(\textbf{r})$ is the total electric field inside the nanoparticle at different position. $\epsilon_0$ is the vacuum permittivity; $\epsilon_{Si}$ is the relative dielectric permittivity of the Si particle. $c$ is the light speed in vacuum; $\hat{U}$ is the 3$\times$3 unity tensor; \textbf{p}, \textbf{T}, \textbf{m}, \textbf{Q} and \textbf{M} are the moments of ED, TD, MD, electric quadrupole (EQ) and MQ respectively. 

\begin{figure}[htbp]
\centering
\includegraphics[width=\linewidth]{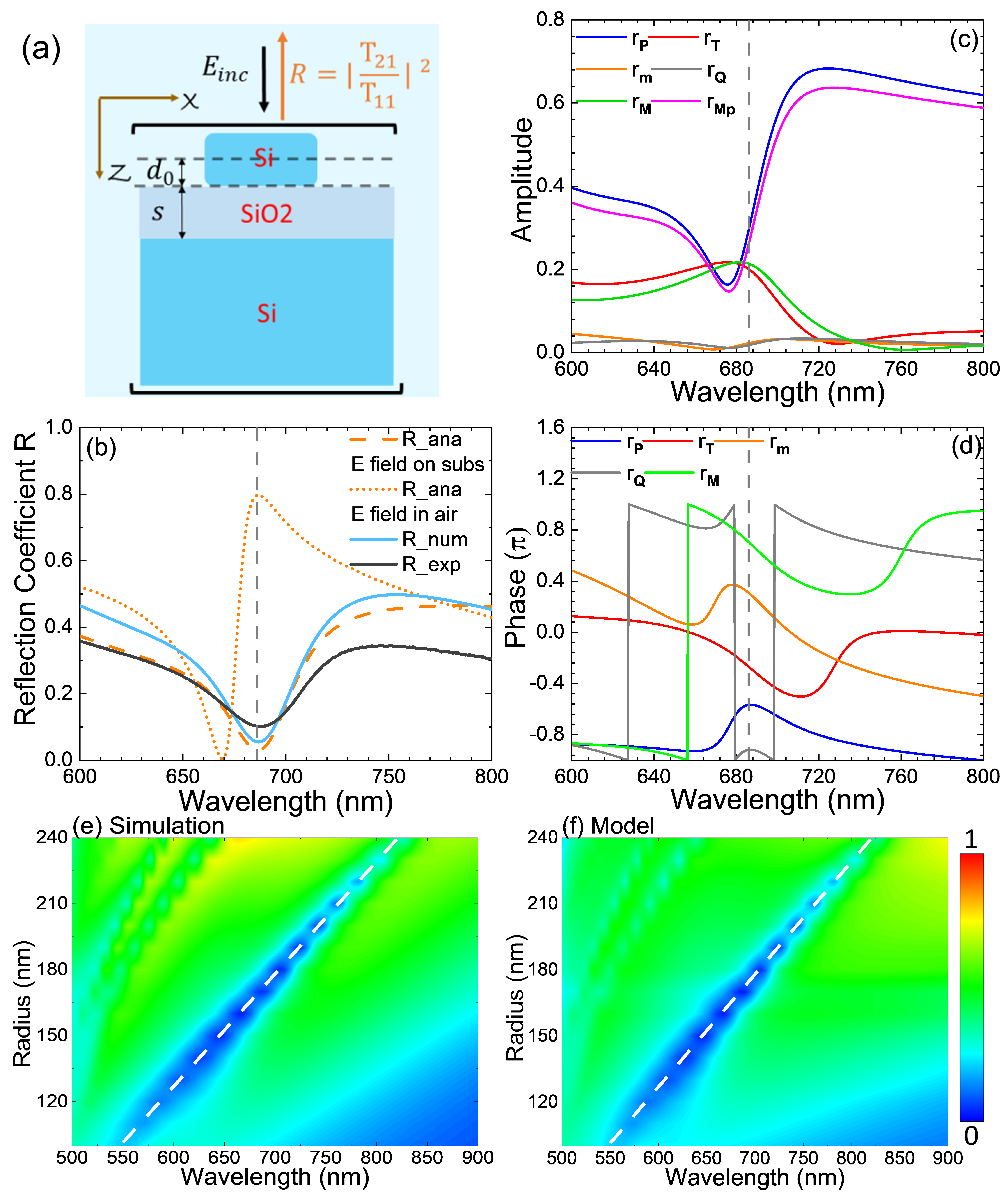}
\caption{(a) Schematic of the incident and reflected fields: The metasurface is illuminated by a normally incident plane wave, $\rm E_{inc}$. A transfer matrix method is used to calculate the reflection coefficient, where each layer of metasurface(air gap)/SiO2/Si geometry is treated as in-line optical element.  (b) The numerically simulated, semi-analytically calculated and experimentally measured reflection coefficient, R, for the Si nanodisk metasurface (radius r = 170 nm, height h= 45 nm). The orange dash (dot) line is the calculation by Eq.3 where multipolar scattering is determined by internal electric field distribution of disk on substrate (in air). The grey dashed line indicates the minimum in reflection at 686 nm. (c) Amplitude of the reflection coefficients for the multipole contributions $\rm r_\textbf{p}$, $\rm r_\textbf{T}$, $\rm r_\textbf{m}$, $\rm r_\textbf{Q}$, $\rm r_\textbf{M}$ and the total multipole contribution $\rm r_{Mp}$. (d) Phase of the multipole contributions in the reflection coefficients,  $\rm r_\textbf{p}$, $\rm r_\textbf{T}$, $\rm r_\textbf{m}$, $\rm r_\textbf{Q}$ and $\rm r_\textbf{M}$.(e) Numerically simulation and (f) semi-analytical model of reflection coefficient R by Eq.3 for nanodisks with radius varying from 100 nm to 240 nm with a step 10 nm. The white dash lines are guide-to-the-eye of the reflection dip. }
\label{fig:F2}
\end{figure}

Under linearly polarized, normally incident light, $\mathrm {E_{inc}= E_0} e^{i(kz-\omega t)}\hat{\textbf{x}}$, the effective contributing multipole moments to the far-field scattering are $\textbf{p}_x$, $\textbf{T}_x$, $\textbf{m}_y$, $\textbf{Q}_{xz}$ and $\textbf{M}_{yz}$ respectively \cite{terekhov2019multipole}. $\rm E_0$ is the amplitude of incident electric field, $k_0=\frac{2\pi}{\lambda}$, is the wavevector in the incident medium, in this case air, and $\omega$ is the frequency of the optical field. The total scattered power includes the forward and backward scattering while the reflection spectra represent backward scattering only. The reflection coefficient is defined as  $\rm R=|r|^2$. In the case where only the Si nanodisks embedded in air is considered, the total contribution of the multipole moments to the reflection coefficient, $\rm r_{Mp}$, is expressed as 
\begin{equation}
\mathrm {r_{Mp}}=\frac{ik_0}{2E_0A\epsilon_0}(\textbf{p}_x+ik_0\textbf{T}_x-\frac{\textbf{m}_y}{c}+\frac{ik_0}{6}\textbf{Q}_{xz}-\frac{ik_0}{2c}\textbf{M}_{yz})
\end{equation}
where $\rm A$ is the area of one metasurface unit cell \cite{terekhov2019multipole}. The contributions from multipole moments to the reflection coefficient $\rm r_{Mp}=r_\textbf{p}+r_\textbf{T}+r_\textbf{m}+r_\textbf{Q}+r_\textbf{M}$, can be expressed as  $\mathrm { r_\textbf{\textbf{p}}}=ik_0\textbf{p}_x/(2E_0A\epsilon_0)$, $\mathrm {r_\textbf{\textbf{T}}}=-k_0^2\textbf{T}_x/(2E_0A\epsilon_0)$, $ \mathrm {r_\textbf{\textbf{m}}}=-ik_0\textbf{m}_y/(2cE_0A\epsilon_0)$, $\mathrm {r_\textbf{\textbf{Q}}}=-k_0^2\textbf{Q}_{xz}/(12E_0A\epsilon_0)$ and $\mathrm {r_\textbf{\textbf{M}}}=k_0^2\textbf{M}_{yz}/(4cE_0A\epsilon_0)$ respectively. 

The amplitude and phase of each multipole mode determines how they interfere with each other and contribute to the far-field scattering. However, the practical metasurface studied here is on a SiO2/Si substrate, the substrate alters the induced electric field distribution within the nanodisk. Therefore, the reflection  transmission coefficient of nanodisk is determined by the integration of the internal electric field within in the particle  sitting on the substrate \cite{bakker2015magnetic, tian2019active}. Additionally, since the studied geometry is a stratified layer, including metasurface/air gap/SiO2 buffer layer/Si layers, we followed the transfer matrix calculations proposed in  Ref. \cite{babicheva2017reflection, chen2012interference} to probe the physics behind the reflection dip, which takes into account Fabry-Perot type multiple reflections within each layer. In the transfer matrix formalism, each layer is considered as a decoupled in-line optical element, where the light interaction is specified by its reflection coefficient  $\rm r$ and transmission coefficient, $\rm t$. The reflection coefficient of overall structure is determined as $\rm R={|\frac{T_{21}}{T_{11}}|}^2$, where $\rm T_{11}$ and $\rm T_{21}$ are the matrix elements in the overall transfer matrix  $\rm{T}$ with $\rm {T =T_{Mp}T_{p1}T_{i1}T_{p2}T_{i2}}$=
$\begin{bmatrix}
\rm T_{11} & \rm T_{12}\\
\rm T_{21} & \rm T_{22}
\end{bmatrix}$.

The transfer matrix of the metasurface-layer with gaps is denoted as $\rm T_{Mp}=\frac{1}{t_{Mp}}
\begin{bmatrix}
1 & \rm{-r_{Mp}}\\
\rm{r_{Mp}} & \rm{t_{Mp}^2-r_{Mp}^2}
\end{bmatrix}$ and $\rm T_{p1}=\begin{bmatrix}
\rm{e^{ik_0d_0}} & 0\\
0 & \rm{e^{-ik_0d_0}}
\end{bmatrix}$, where $\rm{d_0}$ is half the height of the nanodisk. $\rm{t_{Mp}}$ is the transmission coefficient of metasurface and is determined by  
$\rm t_{Mp}=1+\frac{ik_0}{2E_0A\epsilon_0}(\textbf{p}_x+ik_0\textbf{T}_x+\frac{\textbf{m}_y}{c}-\frac{ik_0}{6}\textbf{Q}_{xz}-\frac{ik_0}{2c}\textbf{M}_{yz})$. $\rm{k_b=2\pi\sqrt{\epsilon_{SiO2}}/\lambda}$ is the wavevector in the buffer layer. $\rm s$ is the thickness of SiO2 buffer layer. The transfer matrix of SiO2 buffer layer is denoted as $\mathrm{T_{i1}=\frac{1}{t_{
i1}}\begin{bmatrix}
1 & \rm{r_{i1}}\\
\rm{r_{i1}} & 1
\end{bmatrix}}$ and
$\mathrm{T_{p2}=\begin{bmatrix}
\rm{e^{ik_bs}} & 0\\
0 & \rm{e^{-ik_bs}}
\end{bmatrix}}$, where $\mathrm {r_{i1}=(1-\sqrt{\epsilon_{SiO2}})/(1+\sqrt{\epsilon_{SiO2}})}$,
$\mathrm {t_{i1}=2\epsilon_{SiO2}^{1/4}/(1+\sqrt{\epsilon_{SiO2}})}$. 
The transfer matrix of Si substrate is denoted as
$\mathrm{T_{i2}=\frac{1}{t_{
i2}}\begin{bmatrix}
1 & \rm{r_{i2}}\\
\rm{r_{i2}} & 1
\end{bmatrix}}$, where $\mathrm {r_{i2}=(\sqrt{\epsilon_{SiO2}}-\sqrt{\epsilon_{Si}})/(\sqrt{\epsilon_{SiO2}}+\sqrt{\epsilon_{Si}})}$ and
$\mathrm {t_{i2}=2\epsilon_{SiO2}^{1/4}\epsilon_{Si}^{1/4}/(\sqrt{\epsilon_{SiO2}}+\sqrt{\epsilon_{Si}})}$. $\mathrm{k_0=2\pi/\lambda}$ and $\mathrm {k_b=2\pi\sqrt{\epsilon_{SiO2}}/\lambda}$ are the wavevectors in air and in buffer layer respectively and $\rm{s}$ is the thickness of the SiO2 buffer. 

The semi-analytical expression for the reflection coefficient $\rm R$ of the whole structure becomes
\begin{widetext}
\begin{equation}
\mathrm{R=
|\frac{r_{Mp}(e^{i\varphi_1}+r_{i1}r_{i2}e^{i\varphi_2})+(t_{Mp}^2-r_{Mp}^2)(r_{i1}e^{-i\varphi_2}+r_{i2}e^{-i\varphi_1})}{e^{i\varphi_1}+r_{i1}r_{i2}e^{i\varphi_2}-r_{Mp}(r_{i1}e^{-i\varphi_2}+r_{i2}e^{-i\varphi_1})}|^2}%
\end{equation}
\end{widetext}
where $\mathrm{\varphi_1=-(k_0d_0+k_bs)}$ and $\mathrm{\varphi_2=-(k_0d_0-k_bs)}$ respectively.

The reflection signal is a superposition of several contributing EM waves, including the effective EM multipolar scattered wave generated  by nanodisk and also light reflected  from the SiO2/Si substrate. The light reflected from SiO2/Si substrate  includes the contributions from the incident EM wave directly reflected from the SiO2 and Si substrates, together with Fabry-Perot type multiple reflections due the interfaces within the structure of both the incident and scattered light \cite{babicheva2017reflection}. A schematic can be seen in Fig. \ref{fig:F2} (a), where the reflection coefficient is calculated by the total transfer matrix, which takes into account the multiple reflections as well as the multipolar contribution determined by the induced internal electric field distribution of the nanodisk sitting on the substrate.

The results for the numerical simulation and semi-analytical calculation of the reflection coefficient, $\rm R$, are shown in Fig. \ref{fig:F2} (b). The semi-analytical reflection coefficient, $\rm{R}$, is shown for two cases; i) the internal electric field distribution in the nanodisk for the multipole analysis, and subsequent determination of $\rm r_{Mp}$ and $\rm t_{Mp}$, has been calculated for the nanodisks embedded in air and ii) the electric field distribution is computed for the nanodisks sitting on the SiO2/Si substrate. The semi-analytical model for case (ii) and the numerically calculated reflection spectrum are in excellent agreement with each other and the experimentally measured reflection spectrum, which is also shown in Fig. \ref{fig:F2} (b).

The spectral broadening of experimental measurement results from imperfect fabrication and non-perfect normal incidence in the measurement. A further comparison of the numerical simulations and semi-analytical calculations is presented in Figs. \ref{fig:F2} (e) and (f). The semi-analytical reflection spectra show the same trend and features as the full numerical calculations over a wide range of disk radii from 100 nm to 240 nm. The amplitude and phase of all multipole contributions to the electric field reflection coefficient are shown in Fig. \ref{fig:F2} (c) and (d), respectively. The dominant contributing multipole modes are the ED, TD and MQ. The contributions from the MD and EQ modes are significantly smaller. The reflection minimum at 686 nm is indicated by the grey dash lines in (b)(c) and (d) for more detailed inspection. It is worth noting that the minimum in reflection in Fig. \ref{fig:F2} (b) is blue shifted with respect to the minimum in the amplitude of the reflection coefficient contribution from multipole modes, $\rm r_{Mp}$, seen in Fig. \ref{fig:F2} (c). This indicates the importance of the role played by the substrate. 

The  back-scattered power associated with each multipole moment can be expressed as  $\rm |A|^2$, where A is $\rm r_\textbf{p}$, $\rm r_\textbf{T}$, $\rm r_\textbf{m}$, $\rm r_\textbf{Q}$, or $\rm r_\textbf{M}$. 
However, as mentioned earlier the total reflected power depends on the interference of each component of the field, taking account of the amplitude and relative phase of the scattered fields. If we considered two contributions, A and B, then the sum of their powers yields $\rm I_0$ while taking account of the coherent superposition yields $\rm I_{int}$.
\begin{equation}
\begin{split}
&  \mathrm {I_0=I_A+I_B=|A|^2+|B|^2}
\\&  \mathrm {I_{int}=|A+B|^2=|A|^2+|B|^2+2|A||B|}cos(\varphi_A-\varphi_B)
\end{split}
\end{equation}

Under the circumstance that $\rm I_{int}>I_0$, constructive interference is occurring between the two contributions while destructive interference exists when $\rm I_{int}<I_0$. This is a general criterion to determine constructive or destructive interference and can be applied to any two multipole modes including even higher order modes.

\begin{figure}[htbp]
\centering
\includegraphics[width=\linewidth]{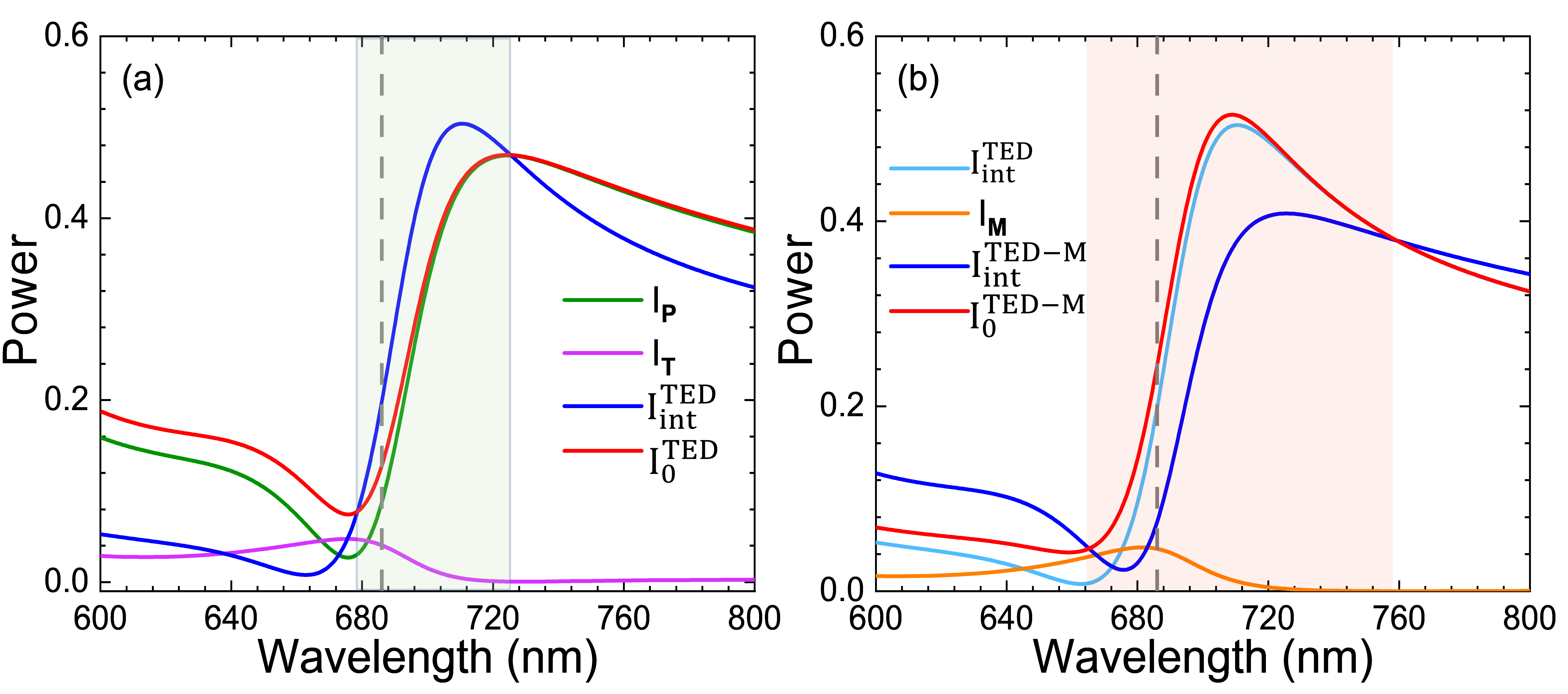}
\caption{(a) The back-scattered power of the ED and TD, $\rm I_P$ and $\rm I_T$ respectively, as well as the sum of their powers, $\rm I_0^{TED}$ and the power of TED, $\rm I_{int}^{TED}$, due to the coherent superposition of the ED and TD contributions. The shaded area indicates the wavelengths at which constructive interference occurs between ED and TD contributions. (b) The back-scattered power of the TED and MQ, $\rm I_{int}^{TED}$ and $\rm I_M$ respectively, as well as the sum of their powers, $\rm I_{0}^{TED-M}$ and the power, $\rm I_{int}^{TED-M}$, due to the coherent superposition of the TED and MQ contributions. The shaded area indicates the wavelengths at which destructive interference occurs between TED and MQ.}
\label{fig:F3}
\end{figure}

As mentioned earlier, the contributions from the ED and TD can form a single contribution called the total electric dipole moment  \cite{terekhov2017multipolar}, or TED$\rm =r_\textbf{p}+r_\textbf{T}$. In order to explore how the ED and TD scattered fields interfere with each other, $\rm I_0^{TED}$ and $\rm I_{int}^{TED}$ are calculated using Eq.4 and are presented in Fig. \ref{fig:F3} (a). It is clear that at the reflection minimum at 686 nm $\rm I_{int}^{TED}>I_0^{TED}$. This demonstrates that there is constructive interference between ED and TD contributions and the reflection dip does not result from an anapole-like mode generated by the ED and TD but rather the TED behaves as a super-dipole at this wavelength and TED results in an increase of the far-field backward scattering. Additionally, it can be seen that there is a wide wavelength range over which there is constructive interference between the ED and TD contribution to the back-scattered radiation. This range is represented by the shaded area in Fig. \ref{fig:F3} (a).

As seen in Fig. \ref{fig:F2}(c), the dominant multipole modes in the studied metasurface are ED, TD and MQ. The multipole reflection coefficient  $\rm r_{Mp}$ can be approximated as  $\rm r_{Mp}\approx r_{TED}+r_\textbf{M}$. The scattering power due to the TED and MQ modes can be seen in  Fig. \ref{fig:F3} (b). It is seen that   $\rm I_{int}^{TED-M}<I_0^{TED-M}$ at 686 nm and there is destructive interference between TED and MD contributions. The wavelengths at which destructive interference exists are presented as the shaded area.  Additionally, $\rm I_{int}^{TED-M}\neq0$ indicates there is only partial destructive interference and the back-scattered field due to the MQ could not totally compensate for the contribution from the TED but results in a decrease of total backward scattering power. Asymmetric scattering manipulation due to the ED and MD interference has been studied for almost 40 years \cite{kerker1983electromagnetic}. Due to the  inherent phase symmetry properties  of MD and MQ modes in the forward and backward directions, where the MD displays odd phase symmetry and the MQ displays an even phase symmetry \cite{liu2019lattice}\cite{liu2018generalized}, the realization of dominant TED and MQ interference in this Letter paves the way towards symmetric scattering manipulation by employing only TED and MQ, such as the unusual transverse scattering, which simultaneously displays zero forward and zero backward scattering for the nanoresonator embedded in an isotropic medium \cite{shamkhi2019transverse}. 
Additionally, our study clearly demonstrated that the substrate effect is nontrivial in reshaping the far-field radiation pattern and must be taken into account for scattering pattern manipulation.

\section{Conclusion}

To conclude, we designed and fabricated an ultra-thin Si nanodisk metasurface on a CMOS-compatible semiconductor-on-insulator wafer. 
Kerker-type suppressed backward scattering is observed through a minimum in the reflection spectrum and is seen to gradually red-shift with increasing disk radius. There is excellent agreement between the numerical simulation and experimental spectra. 
To probe the mechanism producing the suppressed backward scattering, multipole decomposition is performed. A semi-analytical model of backward scattering or reflection coefficient, taking into account the substrate- reflection contribution as well as the multipolar modes' back-scattering is used and found to correspond well with experimental reflection spectrum. Multipole decomposition shows that the dominant modes in the designed Si nanodisk metasurface are the ED, TD and MQ with suppressed MD mode. 
This provided a platform to investigate the Kerker-type scattering due to the ED, TD and MQ rather than conventional ED and MD interference. To address the fundamental issue of constructive or destructive interference of multipolar modes, a generalized criterion employing the amplitude, phase and power analysis is proposed. It is demonstrated that the observed suppressed back-scattering is mainly driven by the coherent interplay between the ED, TD and MQ contribution. The ED back-scattered radiation constructively interferes with that from the TD at the reflection minimum wavelength and the TED behaves as a super-dipole. The suppressed back scattered radiation is realized through destructive interference between the TED and MQ mode contributions. Due to distinct phase symmetry properties of MQ and MD in the forward and backward direction, our findings on TED-MQ interference will spur new Huygens source design or metasurface design for symmetric scattering tailoring, such as transverse scattering. 

\begin{acknowledgments}
We wish to acknowledge the support of Science Foundation Ireland (SFI) under Grant Numbers 16/IA/4550 and 17/NSFC/4918.
\end{acknowledgments}



\nocite{*}

\bibliography{apssamp}

\end{document}